\documentclass[a4paper,aps,pra,preprintnumbers]{revtex4}
\usepackage[utf8]{inputenc}
\usepackage[english,russian]{babel}
\usepackage[T2A]{fontenc}
\usepackage{amssymb,amsfonts,amsmath,mathtext,enumerate,float,dsfont}
\usepackage{graphics,graphicx,epsfig,epstopdf} 
\usepackage{caption}
\usepackage{cmap}    
\usepackage{indentfirst}
\usepackage[usenames]{color}
\usepackage{amsthm}

\begin{document}
\title{CZ vs Beam Splitter: errors of entangled operations}
\author{Zinatullin E. R.}
\author{Korolev S. B.}
\author{Golubeva T. Yu.}
\affiliation{St. Petersburg State University, Universitetskaya nab. 7/9, St. Petersburg, 199034, Russia}
\begin{abstract}

Using the quantum teleportation in continuous variables as a test scheme, we compare two entangled transformations - mixing of signals on a beamsplitter and by the CZ operation. We evaluate these transformations in terms of the errors added to  teleported oscillators.
We have shown that the CZ operation leads to the lower error of teleportation. This error can be further reduced by choosing appropriate weight coefficients for the CZ transforms. We have compared the errors of the theoretical CZ scheme and its practical implementation in the optical design. Although the CZ optical scheme adds intrinsic noise to the overall transformation, it is nevertheless possible to specify the parameters  which provides a gain in comparison with the traditional teleportation protocol. 

\end{abstract}
\maketitle

\section{Introduction}

A teleportation scheme of quantum states is one of the basic protocols known to every student specializing in quantum optics \cite{Bennett,Vaidman,Bouwmeester,Braunstein,Furusawa}. In this paper, we will consider this scheme in the continuous variables regime \cite{Lloyd,Braunstein2}. We want to use this protocol as a test one for comparing two different entangling operations: entangling of squeezed quantum states on a beam splitter and сontrolled-Z (CZ) gate. Both operations are basic transformation tools of linear optics \cite{Larsen,Loock,Alexander1,Su,Alexander2}. Any protocols for quantum computation operates with these elements. The CZ gate is the primary entangling mechanism for generating cluster states in continuous variables \cite{Zhang,Menicucci}, as well as in one-way computing schemes on cluster states \cite{Menicucci,Raussendorf,Nielsen}. It should also be noted that the mechanism of one-way computing itself is based on the teleportation scheme of quantum states \cite{Raussendorf,Menicucci}.

Our attention was attracted by the fact that if in the "classical"\; scheme of quantum teleportation, we replace the beamsplitter that entangles two squeezed states of light with a device that performs the CZ transformation, then the teleportation error associated with the finite squeezing of the initial resource will be lower. This elementary and demonstrative construction posed some questions for us: does the CZ gate always outperform the beamsplitter transformation in terms of added noise? Do the real CZ realizations have the same advantage as a formal mathematical procedure? To what extent is it possible to reduce teleportation noise using CZ gate devices instead of beamsplitters? And finally, what is the reason for the noise decreasing/increasing when one apply one or another entangling operation? In this paper, we will try to answer these questions. 

Besides, the presented consideration allows us to discuss such an aspect of entanglement as weight coefficients \cite{Zhang,Larsen}. It is often assumed by default that unity weight coefficients provide the best entanglement in the system. We will show that variation of this parameter is another opportunity to reduce the noise in the teleportation scheme.

We start with the well-known continuous variable teleportation scheme (Sec. \ref{tp}), then we compare the result with what is obtained by replacing the beamsplitters with entangling CZ gates, bearing in mind arbitrary values of the weight coefficients. We will evaluate the errors of the two schemes and, the best parameters that minimize the teleportation errors will be selected. Finally, in Section \ref{rcz}, we will discuss one of the realizations of the CZ gate and see how the teleportation error will change when we apply this scheme.

\section{Teleportation protocol and its modification using the CZ gate }\label{tp}

\begin{figure}[H]
\includegraphics[scale=1]{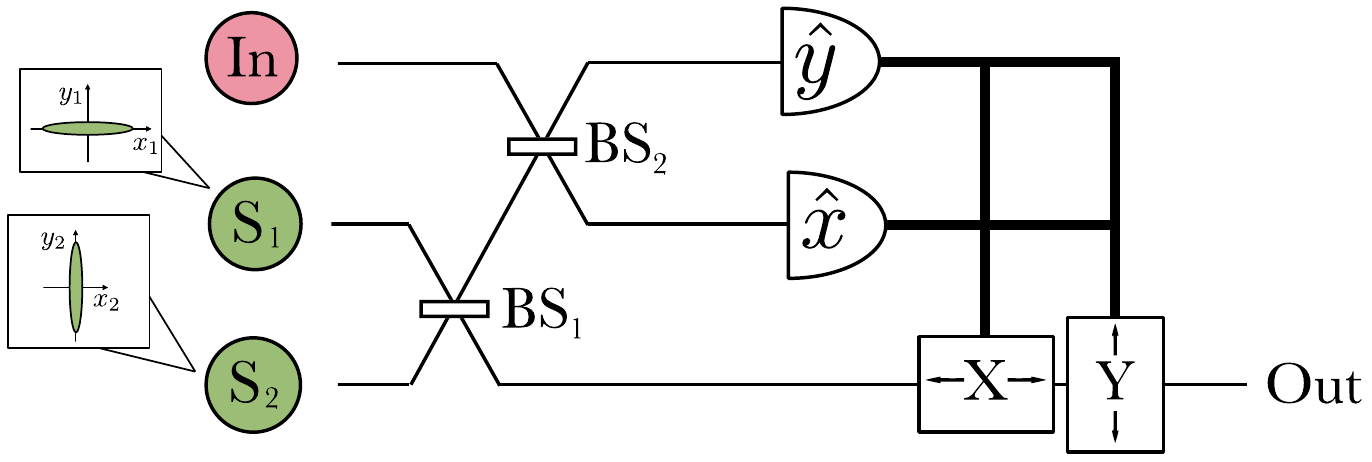}
\centering
\caption{The scheme of teleportation. \textit{In} is input (teleported) state; $S_1$ and $S_2$ are oscillators squeezed in orthogonal quadratures, $\text{BS}_1$ and $\text{BS}_2$ are beamsplitters, $X$ and $Y$ denote operations that shift the corresponding field quadratures.} \label{Fig_0}
\end{figure}

We will start by recalling how the teleportation protocol in continuous variables works. The scheme is shown in Fig. \ref{Fig_0}.
Two oscillators $\text{S}_1$ and $\text{S}_2$ are squeezed in orthogonal directions and are described by quadrature components:
\begin{align}
& \hat{x}_{s,1}=e^r \hat x_{0,1},\qquad\hat{y}_{s,1}=e^{-r} \hat y_{0,1},\label{s1}\\ 
& \hat{x}_{s,2}=e^{-r} \hat x_{0,2},\qquad\hat{y}_{s,2}=e^{r} \hat y_{0,2},\label{s2}
\end{align} 
where $\hat x_{0,j}$ and $\hat y_{0,j}$ are the quadratures of the vacuum state. The squeezing degree is assumed to be equal and is specified by the parameter $r$, which determines the proportional stretch and squeeze of the vacuum field quadratures. The squeezed fields are mixed on the symmetric beamsplitter $\text{BS}_1$, resulting in the entangled state:
 \begin{align}
&\hat{a}_1^\prime = \frac{1}{\sqrt{2}}\left(\left(\hat{x}_{s,1}+\hat{x}_{s,2}\right)+i\left(\hat{y}_{s,1}+\hat{y}_{s,2}\right)\right),\\
& \hat{a}_2^\prime = \frac{1}{\sqrt{2}}\left(\left(\hat{x}_{s,1}-\hat{x}_{s,2}\right)+i\left(\hat{y}_{s,1}-\hat{y}_{s,2}\right)\right).
 \end{align}
This entangled state acts as a quantum resource for teleportation.

Then, the input (teleported) state is added to the field in the first channel (see Fig. 1) with the help of the symmetric beamsplitter $\text{BS}_2$. As a result, the slow field amplitudes after this beamsplitter take the form:
\begin{align}
&\hat{a}_{in}^\prime = \frac{1}{\sqrt{2}}\left(\left(\hat{x}_{in}+  \frac{1}{\sqrt{2}}\left(\hat{x}_{s,1}+ \hat{x}_{s,2} \right)\right)+i\left(\hat{y}_{in}+  \frac{1}{\sqrt{2}}\left(\hat{y}_{s,1}+\hat{y}_{s,2}\right)\right)\right),\\
& \hat{a}_{1}^{\prime\prime} = \frac{1}{\sqrt{2}}\left(\left(\hat{x}_{in}-  \frac{1}{\sqrt{2}}\left(\hat{x}_{s,1}+ \hat{x}_{s,2} \right)\right)+i\left(\hat{y}_{in}-  \frac{1}{\sqrt{2}}\left(\hat{y}_{s,1}+\hat{y}_{s,2}\right)\right)\right).
\end{align}
Then we measure the $y$-quadrature in the input channel and the $x$-quadrature in the first one using balanced homodyne detectors:
 \begin{align}
 \begin{cases}
 \hat{i}_{in,y}=\displaystyle\frac{1}{\sqrt{2}}\beta\left(\hat{y}_{in}+  \frac{1}{\sqrt{2}}\left(\hat{y}_{s,1}+\hat{y}_{s,2}\right)\right),\\
 \hat{i}_{1,x}=\displaystyle\frac{1}{\sqrt{2}}\beta\left(\hat{x}_{in}-  \frac{1}{\sqrt{2}}\left(\hat{x}_{s,1}+ \hat{x}_{s,2} \right)\right).
 \end{cases}
 \end{align}
Here $\beta$ is the homodyne amplitude. Due to the entanglement of the resource state, this measurement will lead to a change of the field quadratures in the second channel:
\begin{align}
& \hat x_2^\prime = \hat x_{in} - \sqrt{2}\hat x_{s,2}- \sqrt{2}i_{1,x}/\beta,\label{8}\\
& \hat y_2^\prime = \hat y_{in} + \sqrt{2}\hat y_{s,1}- \sqrt{2}i_{in,y}/\beta.\label{9}
\end{align}
Here the photocurrent operators are replaced by the c-numbers corresponding to the results of the given measurement.

Finally, the quadratures in the second channel should be shifted on the values of the measured photocurrents to compensate the c-number terms in Eqs. (\ref{8})-(\ref{9}). The state at output of the scheme can be written as:
\begin{align}
& \hat x_{out} = \hat x_{in} - \sqrt{2}\hat x_{s,2} = \hat x_{in} - \sqrt{2}e^{-r} \hat x_{0,2},\label{10}\\
& \hat y_{out} = \hat y_{in} + \sqrt{2}\hat y_{s,1} = 
 \hat y_{in} + \sqrt{2}e^{-r} \hat y_{0,1},\label{11}
\end{align}
where the second equalities take into account the squeezing of the initial resource (\ref{s1})-(\ref{s2}).

Let us now modify the original teleportation scheme by replacing the two beamsplitters with two CZ gates (see Fig. \ref{fig2}) that act as following:
\begin{figure}
\centering
\includegraphics[scale=1]{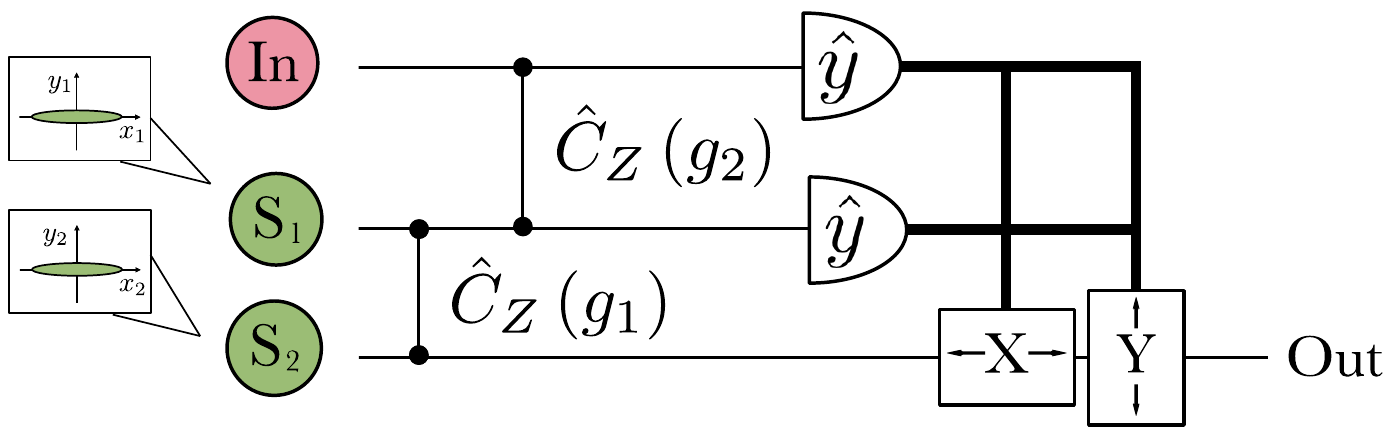}
\caption{The scheme of teleportation with two CZ gates. $\text{S}_1$ and $\text{S}_2$ are oscillators squeezed in $y$-quadratures, $\hat{C}_{Z}(g_i)$ are CZ gates with weight coefficients $g_i$.}
\label{fig2}
\end{figure}
\begin{align}
\hat{C}_{Z,jk}=e^{2ig_{jk}\hat{x}_j\hat{x}_k}.
\end{align}
Here, $g_{jk}$ is the so called weight coefficient of transformation \cite{Zhang,Larsen}. It can take any positive or negative values. Below we will discuss how to choose and control the weight coefficients. For sake of generality, let us assume now that gates have different weight coefficients, correspondingly $g_1$ and $g_2$. 

We now repeat the above derivations  for the modified scheme. Unlike the previous case, we will start with two oscillators squeezed in $y$-quadratures (as it typically assumed for construction of cluster states with the help of CZ gates):
\begin{align}
& \hat{x}_{s,1}=e^r \hat x_{0,1},\qquad\hat{y}_{s,1}=e^{-r} \hat y_{0,1},\label{ss1}\\ 
& \hat{x}_{s,2}=e^{r} \hat x_{0,2},\qquad\hat{y}_{s,2}=e^{-r} \hat y_{0,2}.\label{ss2}
\end{align}
Applying the first CZ gate, we entangle the two squeezed states. Thus, the amplitudes of the oscillators after this transformation can be written as:
 \begin{align}
&\hat{a}_1^\prime = \hat{x}_{s,1}+i\left(\hat{y}_{s,1}+g_{1}\hat{x}_{s,2}\right),\\
& \hat{a}_2^\prime = \hat{x}_{s,2}+i\left(\hat{y}_{s,2}+g_{1}\hat{x}_{s,1}\right).
 \end{align}
Next, applying the second CZ gate, we mix the oscillator in the first channel with the input state that should be teleported. As a result, we get:
  \begin{align}
& \hat{a}_{in}^{\prime}=\hat{x}_{in}+i\left(\hat{y}_{in}+g_2\hat{x}_{s,1}\right),\\
&\hat{a}_1^{\prime\prime} = \hat{x}_{s,1}+i\left(\hat{y}_{s,1}+g_{1}\hat{x}_{s,2}+\hat{g}_2\hat{x}_{in}\right),\\
& \hat{a}_2^\prime = \hat{x}_{s,2}+i\left(\hat{y}_{s,2}+g_{1}\hat{x}_{s,1}\right) \label{1}.
 \end{align}
Let us now measure the $y$-quadratures of the input and first states:
 \begin{align}
 \begin{cases}
 \hat{i}_{y,in}=\hat{y}_{in}+g_2\hat{x}_{s,1},\\
 \hat{i}_{y,1}=\hat{y}_{s,1}+g_1\hat{x}_{s,2}+g_2\hat{x}_{in}.
 \end{cases}
 \end{align}
After replacement of the photocurrents by the corresponding results of the single measurements, we solve the obtained equations with respect to stretched quadratures $\hat x_{s,1}$ and $\hat x_{s,2}$. Then, substituting the solution into (\ref{1}), we get:
 \begin{align}
&\hat{x}_{out}= -\frac{g_2}{g_1}\hat{x}_{in}-\frac{1}{g_1}\hat{y}_{s,1} =  -\frac{g_2}{g_1}\hat{x}_{in}-\frac{1}{g_1}e^{-r} \hat y_{0,1},\label{23}\\
&\hat{y}_{out}= - \frac{g_1}{g_2} \hat{y}_{in}+\hat{y}_{s,2}= - \frac{g_1}{g_2} \hat{y}_{in}+e^{-r} \hat y_{0,2}.\label{24}
 \end{align}
Here we took into account the displacements of the quadrature components on the measurement results and the Eqs. (\ref{ss1})-(\ref{ss2}). Putting $g_1=-g_2$ (that corresponds to the different phases of two gates), we get the teleportation transformation of the form:
 \begin{align}
&\hat{x}_{out}= \hat{x}_{in}-\frac{1}{g_1}e^{-r} \hat y_{0,1},\\
&\hat{y}_{out}= \hat{y}_{in}+e^{-r} \hat y_{0,2}.
 \end{align}

Let us compare the obtained result with the results of the original teleportation scheme (\ref{10})-(\ref{11}).

We want to emphasize two important facts: 
\begin{enumerate}
\item When the weight coefficient $g_1$ equals unity, the scheme with CZ gates provides a lower error  compared to the original teleportation scheme with two beamsplitters. It is convenient to characterize the error level by the value of the mean square errors in each quadrature $\langle \delta \hat{e}^2_x \rangle= \langle (\hat x_{out} - \hat x_{in})^2\rangle\;,\quad \langle \delta \hat{e}^2_y \rangle= \langle (\hat y_{out} - \hat y_{in})^2\rangle\;$. One can see that the error level determined by Eqs. (\ref{10})-(\ref{11}) is two times higher than for scheme with the CZ gates (supposing $g_1=-g_2=1$). In this case, the errors in $x$ and $y$-quadratures coincide.
\item The second scheme (with CZ) allow us to further reduce the errors in one of the quadratures, in principle, admit of the reach of zero error at $g_1=-g_2 \longrightarrow \infty$.
\end{enumerate}

Thus, the obtained results indicate a significant advantage of using the CZ gates over the entanglement of the fields with the beamsplitters. However, talking about the CZ gate, we assumed some abstract device that performs the necessary operation, while the beam splitter is a specific implementation of the entangling procedure. In this regard, it would be correct to compare the available CZ gate realisations with the results that the beamsplitter ensures. We will perform such a comparison in Section \ref{rcz}, but for now, consider another teleportation scheme, in which the squeezed fields are entangled using the CZ gate, and the input state of light is mixed using the beamsplitter.
  
\section{Teleportation in a hybrid scheme} \label{htp}

\begin{figure}[H]
 \centering
 \includegraphics[scale=1]{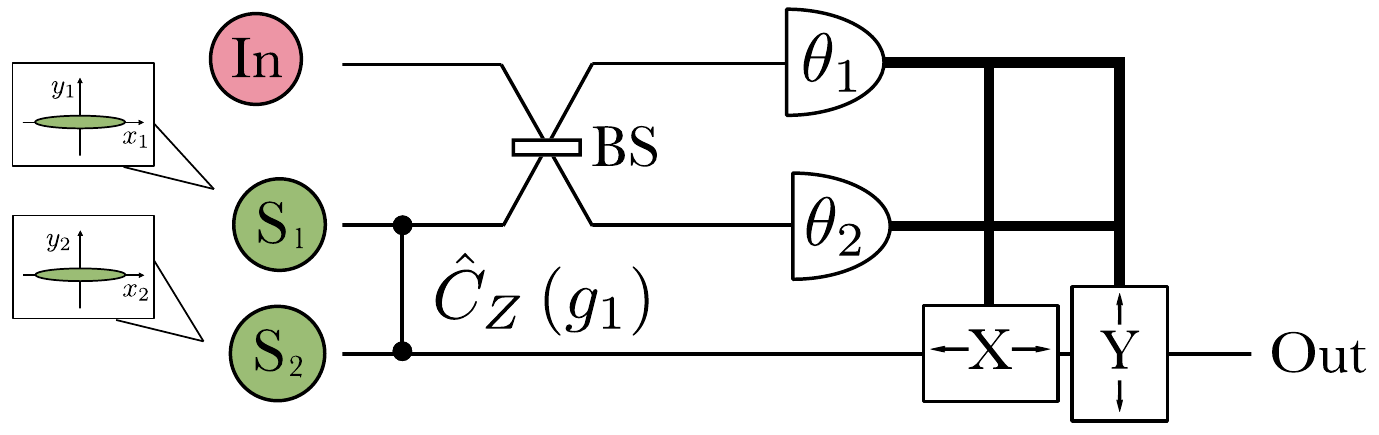}
 \caption{The scheme of teleportation with both the CZ gate and the beamsplitter. $\theta _1$ and $\theta _2$ are the phases of the local oscillators employed for balanced homodyne detection.
} \label{fig3}
 \end{figure}

Let us consider a hybrid teleportation scheme where both a beamsplitter and a CZ gate are present. We suppose, the entangling of squeezed fields is carried out by the CZ gate, and the input state of light is admixed by the symmetric beamsplitter (see Fig. \ref{fig3}).
Carrying out derivations similar to those performed in the previous section, we find that in this case, the quadratures are transformed by the way:
 \begin{align}
&\hat{a}_{in}^{\prime} =\frac{1}{\sqrt{2}}\left(\hat{x}_{in}+\hat{x}_{s,1}\right)+\frac{i}{\sqrt{2}}\left(\hat{y}_{in}+\hat{y}_{s,1}+g_1\hat{x}_{s,2}\right),\\
&\hat{a}_{1}^{\prime} =\frac{1}{\sqrt{2}}\left(\hat{x}_{in}-\hat{x}_{s,1}\right)+\frac{i}{\sqrt{2}}\left(\hat{y}_{in}-\hat{y}_{s,1}-g_1\hat{x}_{s,2}\right),\\
&\hat{a}_{2}^{\prime} =\hat{x}_{s,2}+i\left(\hat{y}_{s,2}+g_1\hat{x}_{s,1}\right). \label{27}
 \end{align}
Now we perform homodyne detection in two channels - the input channel and the first one. In the equations below we keep the phases $\theta _1$ and $\theta _2$ of the local oscillators as variable parameters:
 \begin{align}
 \begin{cases}
\sqrt{2} \hat{i}_{in}=\cos \theta_1\left(\hat{x}_{in}+\hat{x}_{s,1}\right)+\cos \theta_1\left(\hat{y}_{in}+\hat{y}_{s,1}+g_1\hat{x}_{s,2}\right),\\
\sqrt{2} \hat{i}_{1}=\cos \theta_2\left(\hat{x}_{in}-\hat{x}_{s,5}\right)+\cos \theta_2\left(\hat{y}_{in}-\hat{y}_{s,5}-g_1\hat{x}_{s,6}\right).
 \end{cases}
 \end{align}
Solving the system with respect to $x$-quadratures and substituting this solution into (\ref{27}), we get:
 \begin{align}
\begin{pmatrix}
\hat{x}_{out}\\
\hat{y}_{out}
\end{pmatrix}=\frac{1}{\sin \theta _-}\begin{pmatrix}
-\frac{\cos \theta _-+\cos \theta_+}{g_1} & -\frac{\sin \theta_+}{g_1}\\
g_1\sin \theta_+ & g_1 \left(	\cos \theta_--\cos \theta _+\right)
\end{pmatrix} \begin{pmatrix}
\hat{x}_{in}\\
\hat{y}_{in}
\end{pmatrix}+\begin{pmatrix}
-\frac{\hat{y}_{s,1}}{g_1}\\
\hat{y}_{s,2}
\end{pmatrix}+\begin{pmatrix}
-\frac{1}{g_1} & \frac{1}{g_1}\\
g_1 & g_1
\end{pmatrix}\begin{pmatrix}
i_{in}\\
i_1
\end{pmatrix},\label{29}
 \end{align}
where $\theta _\pm =\theta_1\pm \theta_2$. Here, as before, the last term in (\ref{29}) is related to the results of the measurements and can be compensated. To convert this transformation into teleportation, we have to choose the appropriate values of $\theta_1$ and $\theta_2$. If we put $\theta_2=-\theta_1=\pi/4$, then the Eq. (\ref{29}) is turned into:
 \begin{align}
 \begin{pmatrix}
\hat{x}_{out}\\
\hat{y}_{out}
\end{pmatrix}=\begin{pmatrix}
\frac{1}{g_1} & 0\\
0 & g_1
\end{pmatrix} \begin{pmatrix}
\hat{x}_{in}\\
\hat{y}_{in}
\end{pmatrix}+\begin{pmatrix}
-\frac{\hat{y}_{s,1}}{g_1}\\
\hat{y}_{s,2}
\end{pmatrix}.\label{30}
 \end{align}
One can easily see that this transformation tunes out to be teleportation only for $g_1=1$. Otherwise, instead of the input state, we obtain a state to which the squeezing transform is applied. Thereby, we can not vary the weight coefficient to reduce the noise in this scheme. Thus, the solution (\ref{30}) will have the same error as in the previous case, only if the weight coefficient in the (\ref{23}) is also put equal to one. It can be concluded that such the hybrid scheme provides a lower error than the configuration with two beamsplitters, but it loses to configuration with two CZ gates.

Let us now move from the idealized CZ gate to its realizations.
 
 \section{ Teleportation with real CZ gate
 }\label{rcz}
 
\begin{figure}[H]
\centering
\includegraphics[scale=1.2]{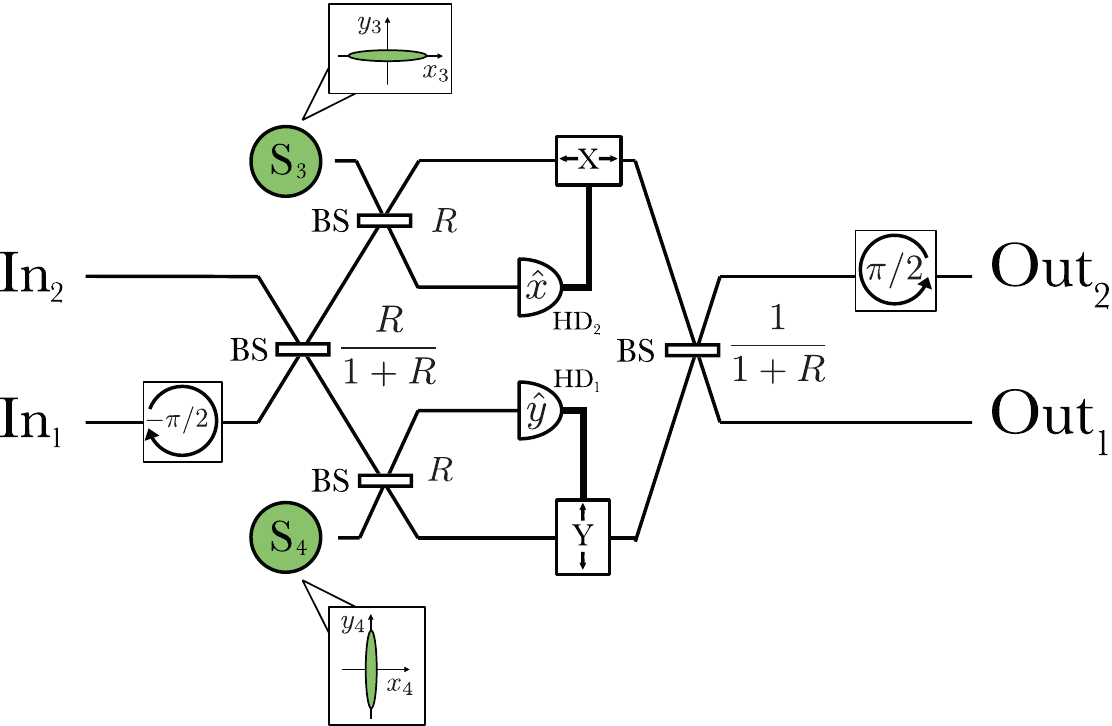}
\caption{Scheme of optical CZ gate realization.}
\label{fig4}
\end{figure}

Up to now, we considered the ideal CZ gate that itself was implemented without errors. This means, the error imposed by the finite squeezing degree of the resource states was the only error that affected on the results in Sections \ref{tp} and \ref{htp}. In reality, CZ gate itself is error prone. Let us follow these errors examining the optical realization of the CZ gate. The circuit have shown in Fig. \ref{fig4}. This scheme almost completely repeats the one proposed in  \cite{realisticCz} for the experimental implementation of the SUM-gate (analogue of CNOT-gate in continuous variables). We slightly modified that scheme by adding two phase shifters at the input of the first channel and at the output of the the second channel, to make it execute the CZ transform.

Let us consider the operation of the optical CZ gate  realization. First, the phase of the first oscillator is rotated on $-\pi/2$ and two input states are mixed on the beamsplitter with a reflection coefficient $R/(1+R)$. Then, auxiliary oscillators in squeezed states are entangled to the states in each of the channels. This mixing carried out on the beamsplitters with the amplitude reflection coefficient $R$. The oscillator in the first channel is entangled with the auxiliary oscillator squeezed in $x$-quadrature, and in the second one - with auxiliary oscillator squeezed in $y$-quadrature. Then, one of the states in each channel is measured by the homodyne detector: the $y$-quadrature is measured in the first channel and the $x$-quadrature -- in the second channel. The quadratures of the remaining oscillators are displaced according to the measurement values. To complete the CZ gate, the states are mixed on the beam splitter with the reflectivity $1/(1+R)$, and the phase of the second state is rotated by $\pi/2$. Since the scheme employs two auxiliary squeezed oscillators and four beam splitters one can expect that these devices will introduce a fair amount of additional noise.

The circuit shown in Fig. \ref{fig4}  implements the following transformation:
\begin{align}
\begin{pmatrix}
\hat{x}_{out,1}\\
\hat{x}_{out,2}\\
\hat{y}_{out,1}\\
\hat{y}_{out,2}
\end{pmatrix}=\begin{pmatrix}
1 & 0 & 0 & 0\\
0 &  1 & 0 & 0 \\
0 & \frac{1-R}{\sqrt{R}} & 1 & 0\\
 \frac{1-R}{\sqrt{R}} & 0 & 0 & 1
\end{pmatrix}\begin{pmatrix}
\hat{x}_{in,1}\\
\hat{x}_{in,2}\\
\hat{y}_{in,1}\\
\hat{y}_{in,2}
\end{pmatrix}+\sqrt{\frac{1-R}{1+R}} 
\begin{pmatrix} 
-\hat{x}_{s,4}\\
\hat{y}_{s,3}\\
-\sqrt{R} \hat{y}_{s,3}\\
-\sqrt{R} \hat{x}_{s,4}
\end{pmatrix}.
\label{31}
\end{align}
Let us check the level of the teleportation error when this CZ device is employed as entangled gate. We will imply that both CZ gates in the scheme are identical. As we have seen above, this condition ensure the equality  of the weight coefficients and the circuit implements a teleportation protocol.

\subsection{Optical CZ gate in the teleportation circuit with two CZ gates}

\begin{figure}[H]
\centering
\includegraphics[scale=1]{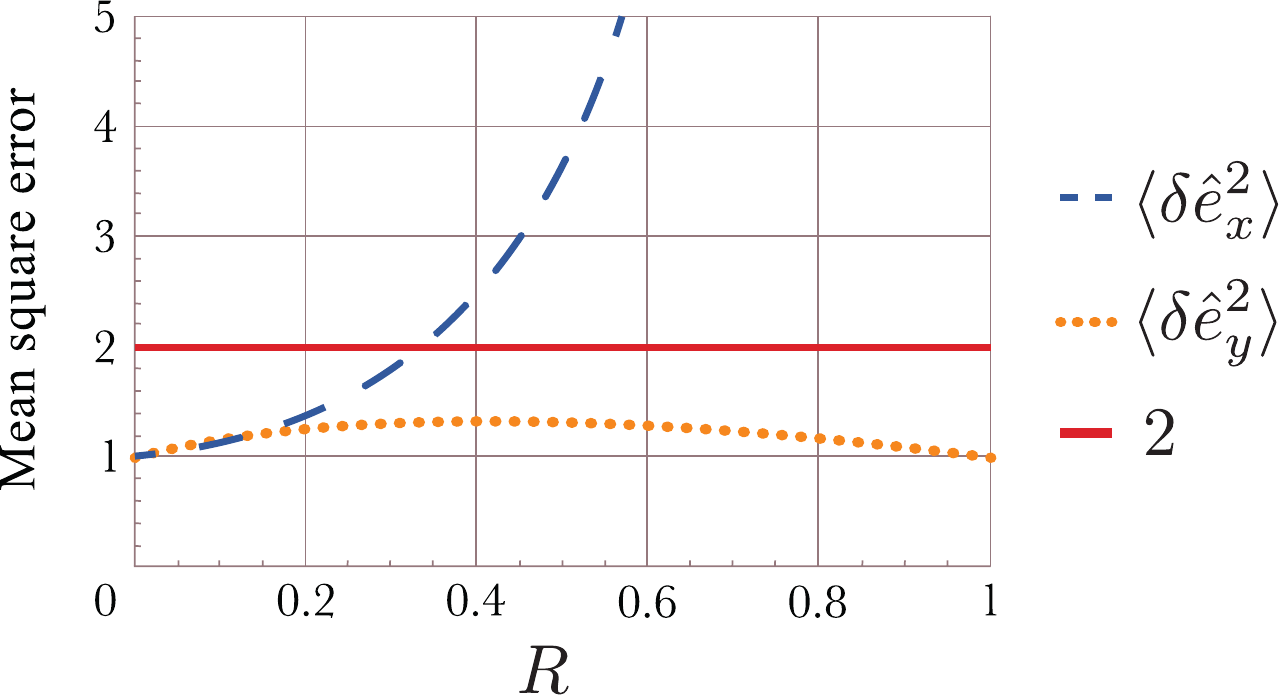}
\caption{Mean square errors of the quadratures of teleported signal, depending on the reflection coefficient of the beamsplitters used in the optical CZ gates. The solid red line indicates the noise level of teleportation in the original scheme with the beamsplitters.}
\label{fig5}
\end{figure}

If we now implement the teleportation with two CZ gates (discussed in Sec. \ref{tp}) using optical CZ transformation (\ref{31}), we get:
\begin{align}
&\hat{x}_{out}=-\hat{x}_{in}-\frac{\sqrt{R}}{1-R}\hat{y}_{s,1}+\frac{R}{\sqrt{1-R^2}}\hat{x}_{s,4}+\frac{1}{\sqrt{1-R^2}}\hat{y}_{s,5},\\
&\hat{y}_{out}=-\hat{y}_{in}+\hat{y}_{s,2}+\sqrt{R}\frac{\sqrt{1-R}}{\sqrt{1+R}}\left(\hat{x}_{s,6}-\hat{y}_{s,3}\right).
\end{align}
Here the first terms correspond to the desired teleportation of the input state. The second terms (with indices 1 and 2) arise because of the finite squeezing of two resource oscillators in the teleportation protocol. The remaining terms (with indices 3,4,5, and 6) are due to auxiliary squeezed oscillators in each of the two CZ gates. Now let us estimate the errors of this teleportation protocol, more precisely,  their mean square fluctuations. Here, as before, we assume squeezed oscillators be statistically independent. Moreover, we suppose that all oscillators are squeezed equally. Taking all of these into account, the following equalities can be derived:
\begin{align}
&\langle \delta \hat{e}^2_x \rangle  =\left(\frac{1+(2-R)R^2}{(1-R)^2(1+R)}\right)\langle\delta \hat{y}_s^2\rangle,\\
&\langle \delta \hat{e}^2_y \rangle=\left(\frac{1+3R-2R^2}{1+R}\right) \langle\delta \hat{y}_s^2\rangle.    
\end{align}
In Fig. \ref{fig5} mean square error for a given squeezing is plotted. For comparison, the solid red line indicates the noise level when teleportation is implemented in the original scheme with two beamsplitters.
First of all, one can see that in this case, the beamsplitter reflection coefficient play the role of the control parameter (that is, determine the value of the weight coefficients of CZ). However, unlike expressions (\ref{23})-(\ref{24}), the errors of both quadratures will rely on this parameter. Moreover, the $x$-quadrature error dependence on the value of $R$ is inverse in comparison with expression (\ref{23}).
 
One can see from the figure that an increase in $R$ leads to the increase in the error of $x$-quadrature, whereas the error in $y$-quadrature is nearly independent of $R$ and does not exceed 2.

Thus, the region of \textbf{$R< 0.33 $} can be revealed, where the optical scheme with CZ gates has an advantage over the original teleportation protocol with beamsplitters, providing a lower error level.

Let us now evaluate the application of the considered optical CZ gate to the hybrid teleportation scheme discussed in Sec. \ref{htp}.

\subsection{Applying the optical CZ gate to the hybrid teleportation scheme}

In this section we will apply the optical CZ transformation (\ref{31}) to the hybrid teleportation scheme discussed in the section \ref{htp}. In this case, the following formulas for the quadratures of teleported signal can be written:
\begin{align}
&\hat{x}_{out}=\frac{\sqrt{R}}{1-R}\hat{x}_{in} -\frac{\sqrt{R}}{1-R}\hat{y}_{s,1}+\frac{1}{\sqrt{1-R^2}}\hat{y}_{s,3},\\
&\hat{y}_{out}=\frac{1-R}{\sqrt{R}}\hat{y}_{in}+\hat{y}_{s,2}+(1-2R)\sqrt{\frac{1-R}{R(1+R)}}\hat{x}_{s,4}.
\end{align}
As before, the only one value of the weight coefficient turns the considered transform into the teleportation. If we put $R=\frac{1}{2}\left(3 - \sqrt {5}\right)\approx 0.38$, then we get
\begin{align}
&\hat{x}_{out}=\hat{x}_{in} -\hat{y}_{s,1}+\sqrt{\frac{1}{10}\left(5 +3 \sqrt{5}\right)}\hat{y}_{s,3},\\
&\hat{y}_{out}=\hat{y}_{in}+\hat{y}_{s,2}+\sqrt{\frac{1}{10}\left(7\sqrt{5}-15\right)}\hat{x}_{s,4}.
\end{align}
Then, mean square errors are given as follows:
\begin{align}
&\langle \delta \hat{e}_x^2\rangle =\frac{3\sqrt{5}}{10}\left(1 +\sqrt{5}\right) \langle \delta \hat{y}_{s} \rangle \approx 2.17 \langle \delta \hat{y}_{s} \rangle, \\
&\langle \delta \hat{e}_y^2\rangle =\frac{\sqrt{5}}{10}\left(7 -\sqrt{5}\right) \langle \delta \hat{y}_{s} \rangle \approx 1.07 \langle \delta \hat{y}_{s} \rangle.
\end{align}
Comparing this result with (\ref{30})  (teleportation with ideal CZ gate), we notice that the situation has changed significantly. In the scheme under consideration the error in $x$-quadrature turns out higher than 2, and the error in $y$-quadrature is lower then 2, whereas in the ideal CZ gate these errors coincide and equal unity. In this case, it cannot be argued clearly the advantage of this protocol over the original one (with two beamsplitters).

\section{Conclusion}

In this paper, we have compared two entangling transformations: CZ gate and mixing signals on the beamsplitter. We tested the quality of these transformations in terms of adding an error under the teleportation protocol. We have shown that employing the CZ gate bring in lower noise, thus providing a higher quality of protocol performance. This conclusion is valid not only for an idealized CZ scheme but also for real one, for example, for optical CZ gate in some range of reflection coefficient values.

We have proven that the choice of the weight coefficients in CZ gate  is a supportive factor in noise control and ideally it allows to reach near the zero error in one of the quadratures.

Using the hybrid teleportation protocol, where the resource state is obtained by the CZ transform, and the input state is admixed on the beamsplitter, makes us denied to vary the weight coefficients, since only one value convert the transfomation into teleportation. The theory predicts that an ideal (no noise-introducing) CZ gate will give a gain in noise compared to the original teleportation scheme, however, the considered optical implementation of the CZ gate cannot be considered as preferable. 

Thus, although teleportation circuits based on CZ transform are more complex than the original beamsplitter configuration, they ensure higher accuracy of teleportation due to reduced errors. This difference related with fixed signal-to-noise ratio imposed by the symmetrical beamsplitter, while the varying of the weight coefficients allows to enlarge this ratio in CZ. Even when we consider a non-ideal CZ transform, there is a range of parameters at which the amount of noise introduced by the auxiliary squeezed states is less than the improvement provided by weight coefficients. Therefore, search for better CZ gates can enhance the accuracy of the teleportation protocol. 

The optical implementation of CZ that we have focused on in this paper is far from the only one; there are quite a few variants of this gate, performed by means of different physical systems \cite{Wang,Leibfried,Song,Wang1,Collodo,Ciccarello,Paredes,Levine}. Among them, we would like to note the schemes based on the interaction of light beams inside the atomic ensemble \cite{Wang}. In this scheme, the weight coefficient is determined by the pulse durations and, as a result, can be varied easily.
The optimal value of the weight coefficients in each particular scheme can be considered as a balance between the reduction of noise from the resource states and the simultaneous adding of noise from the CZ transform itself. 

Another important problem is the complexity of the experimental implementation of the scheme we have discussed above. To implement just one CZ gate, we need two additional light sources, a pair of homodyne detectors and four asymmetric beamsplitters, with different but precisely matched reflectances. In the assessments above, we have estimated only the unavoidable noise associated with the finite squeezing degree of the oscillators used both as a resource and to create the CZ gate. However, any deviation of this scheme from the exact balance may deprive us of the discussed advantage. For other existing implementations of the CZ \cite{Wang,Leibfried,Song,Wang1,Collodo,Ciccarello,Paredes,Levine}, the situation is not much better. At the same time, the original teleportation scheme is much simpler and less demanding on experimental equipment. 

It should be noted that the issue of increasing the fidelity of teleportation can be solved by using non-Gaussian states as a resource \cite{Opatrny,Cochrane}. To obtain such states, the procedure of conditional subtraction or addition of photons is used. We propose here a method that allows one to increase the fidelity of teleportation while remaining within the framework of Gaussian deterministic operations.

\vspace{0.5 cm}

This research was supported by the Russian Foundation for Basic Research (RFBR) under the project 19-02-00204a.

\end{document}